\documentclass[twocolumn,aps,superscriptaddress,showpacs]{revtex4}

\usepackage{amssymb}
\usepackage{amsmath}
\usepackage{graphicx}
\usepackage[normalem]{ulem}
\usepackage[dvips]{color}

\begin{document}

\title{Spin-orbit coupling and the up-down differential transverse flow in intermediate-energy heavy-ion collisions}
\author{Yin Xia}
\affiliation{Shanghai Institute of Applied Physics, Chinese Academy
of Sciences, Shanghai 201800, China}
\author{Jun Xu\footnote{corresponding author: xujun@sinap.ac.cn}}
\affiliation{Shanghai Institute of Applied Physics, Chinese Academy
of Sciences, Shanghai 201800, China}
\author{Bao-An Li}
\affiliation{Department of Physics and Astronomy, Texas A$\&$M
University-Commerce, Commerce, TX 75429-3011, USA}
\affiliation{Department of Applied Physics, Xi'an Jiao Tong
University, Xi'an 710049, China}
\author{Wen-Qing Shen}
\affiliation{Shanghai Institute of Applied Physics, Chinese Academy
of Sciences, Shanghai 201800, China}

\date{\today}

\begin{abstract}
To explore the strength, isospin dependence, and density dependence
of the in-medium spin-orbit coupling that is poorly known but
relevant for understanding the structure of rare isotopes, the
nucleon spin up-down differential transverse flow is studied
systematically by varying the beam energy and impact parameter of
Au+Au collisions within a spin-isospin dependent transport model
recently developed for investigating spin-related phenomena in
intermediate-energy heavy-ion collisions. The optimal reaction
conditions for delineating and disentangling effects of different
terms in the spin-orbit coupling are discussed.
\end{abstract}


\maketitle

\section{Introduction}
\label{introduction}

The importance of nuclear spin-orbit interaction was first
recognized after Goeppert-Mayer~\cite{Goe49} and Haxel {\it et
al}~\cite{Hax49} used it to explain successfully the magic numbers
65 years ago. Although in free space the spin-orbit interaction can
be well fitted by the nucleon-nucleon scattering data, the in-medium
spin-orbit interaction, which is known to be critical for
determining properties of drip-line nuclei~\cite{Lal98}, the
astrophysical r-process~\cite{Che95}, and the location of stability
island for superheavy elements~\cite{Ben99,Mor08}, still can not be
well defined. Despite of the extensive studies of the shell
structure and other properties of finite nuclei, there exists major
uncertainties and interesting puzzles regarding the strength,
density dependence, and isospin dependence of the spin-orbit
coupling. For instance, the strength of the spin-orbit coupling has
only been constrained between $80$ and $150$ MeVfm$^5$ by studying
nuclear structures~\cite{Les07,Zal08,Ben09}. While the structure of
"bubble nuclei"~\cite{Tod04,Gra09,Sor13} provides an interesting
opportunity for studying the density dependence of the spin-orbit
coupling, the density range and gradient explored are rather small
compared to that reached in nuclear reactions at intermediate
energies. The isospin dependence of the spin-orbit coupling refers
to the relative strength of the isospin-like nucleons with respect
to the isospin-unlike ones. It was found that the kink of the charge
radii of lead isotopes favors a similar coupling strength for the
isospin-like and isospin-unlike nucleons~\cite{Sha95,Rei95}, while
strong experimental evidences of a decreasing spin-orbit coupling
strength with increasing neutron excess were
reported~\cite{Sch04,Gau06}.

The so-call "Spin Hall Effect"~\cite{Hir99} due to the spin-orbit
coupling is expected to be a general phenomenon during particle
transport processes. In nuclear reactions, effects of the spin-orbit
coupling have already been studied at both
low~\cite{Uma86,Mar06,Iwa11} and ultra-relativistic
energies~\cite{Lia05}. To investigate the spin-dependent phenomena
and explore the strength, isospin dependence, and density dependence
of the in-medium spin-orbit interaction, we have recently developed
a spin-isospin dependent transport model SIBUU12~\cite{Xu13} by
incorporating the spin degree of freedom and the spin-dependent
nucleon potentials into the isospin-dependent
Boltzmann-Uehing-Uhlenbeck (IBUU) transport model~\cite{Li97,LCK}
for intermediate-energy heavy-ion collisions. The nucleon spin
up-down differential transverse flow was found to be sensitive to
the variation of parameters for the spin-orbit coupling in Au+Au
reaction at a beam energy of 50 MeV/nucleon and an impact parameter
of $\text{b}=8$ fm~\cite{Xu13}. However, the strength, density
dependence, and isospin dependence will all affect the spin up-down
differential flow, and it is thus difficult to extract clear
information about different components of the spin-orbit coupling
with a single reaction. Compared with nuclear spectroscopy and
corresponding nuclear structure studies, heavy-ion reactions has the
advantage of adjusting the density, beam energy, and momentum flux
of the system, giving more options to study the detailed properties
of the spin-orbit coupling. In this work, we perform a systematic
study of the nucleon spin up-down differential transverse flow by
varying the beam energy and impact parameter of Au+Au collisions.
The optimal reaction conditions and the possibility of delineating
and disentangling effects of different terms in the spin-orbit
coupling by using high transverse momentum nucleons are discussed.

The paper is organized as follows. In Sec.~\ref{model} we outline
the spin-orbit interaction and the spin-dependent mean-field
potentials within Hartree-Fock calculations. In Sec.~\ref{sibuu12},
we describe how the spin degree of freedom and the spin-dependent
potentials are incorporated into the SIBUU12 transport model. In
Sec.~\ref{results}, we present results of our systematic studies on
the spin up-down differential transverse flow and discuss the
possibility of delineating and disentangling effects of the
strength, isospin dependence, and density dependence of the
spin-orbit coupling. Finally, a summary is given at the end.

\section{Spin-orbit interaction and spin-dependent mean-field potentials}
\label{model}

We begin with the Skyrme type effective nuclear spin-orbit
interaction between two nucleons at position $\vec{r}_1$ and
$\vec{r}_2$~\cite{Vau72}\label{vso}
\begin{equation}
V_{so} = i W_0 (\vec{\sigma}_1+\vec{\sigma}_2) \cdot \vec{k} \times
\delta(\vec{r}_1-\vec{r}_2) \vec{k}^\prime,
\end{equation}
where $W_0$ is the strength of the spin-orbit coupling,
$\vec{\sigma}_{1(2)}$ is the Pauli matrix,
$\vec{k}=(\vec{p}_1-\vec{p}_2)/2$ is the relative momentum operator
acting on the right with $\vec{p}=-i\nabla$, and $\vec{k}^\prime$ is
the complex conjugate of $\vec{k}$. Within the framework of
Hartree-Fock calculation, the above nuclear interaction leads to the
time-even and time-odd spin-dependent mean-field potentials written as
\begin{eqnarray}
U_q^{s-even} &=& -\frac{W_0}{2} [\nabla \cdot (\vec{J} + \vec{J}_q)
] \nonumber \\
&+& \frac{W_0}{2}(\nabla \rho + \nabla \rho_q) \cdot (\vec{p} \times
\vec{\sigma}),\label{useven} \\
U_q^{s-odd} &=& - \frac{W_0}{2}\vec{p} \cdot [\nabla \times (\vec{s}
+\vec{s}_q) ] \nonumber \\
&-& \frac{W_0}{2} \vec{\sigma} \cdot [\nabla \times (\vec{j} +
\vec{j}_q) ],\label{usodd}
\end{eqnarray}
where
\begin{eqnarray}
\rho &=& \sum_i \phi^\star_i \phi_i,\\
\vec{s} &=& \sum_i \sum_{\sigma,\sigma^\prime} \phi^\star_i
\langle\sigma|\vec{\sigma}|\sigma^\prime\rangle \phi_i, \\
\vec{j} &=& \frac{1}{2i} \sum_i (\phi^\star_i \nabla\phi_i- \phi_i
\nabla\phi^\star_i),\\
\vec{J} &=& \frac{1}{2i} \sum_i \sum_{\sigma,\sigma^\prime}
(\phi^\star_i \nabla\phi_i- \phi_i \nabla\phi^\star_i)\times
\langle\sigma|\vec{\sigma}|\sigma^\prime\rangle,
\end{eqnarray}
are respectively the number, spin, momentum, and spin-current
densities, with $\phi_i$ being the wave function of the $i$th
nucleon. In Eqs.~(\ref{useven}) and (\ref{usodd}) $q=n$ or $p$ is
the isospin index. It is noteworthy that although only time-even
potentials are considered in the study of spherical nuclei, time-odd
potentials should be included in the study of deformed nuclei. In
heavy-ion collisions with all kinds of geometric asymmetries, both
time-even and time-odd potentials are necessary, otherwise the
Galilean invariance will be broken and the frame-dependent spurious
spin polarization will appear~\cite{Mar06}.

The second term in Eq.~(\ref{useven}) is usually called the
spin-orbit potential written in the form of
$\vec{W}_{q}\cdot(\vec{p} \times \vec{\sigma})$ with the form factor
expressed as
\begin{equation}
\vec{W}_{q} =\frac{W_0}{2}(\nabla \rho + \nabla \rho_q).
\end{equation}
In Ref.~\cite{Pea94} a density-dependent nuclear spin-orbit
interaction is generalized, and the form factor of the spin-orbit
potential has the form of
\begin{eqnarray}
\vec{W}_{q}&=&\frac{W_0}{2}\nabla(\rho+\rho_q)+\frac{W_1}{2}
[\rho^\gamma\nabla(\rho-\rho_{q})\nonumber \\
&+&(2+\gamma)(2\rho_{q})^\gamma\nabla\rho_{q}]+\frac{W_1}{4}\gamma\rho^{\gamma-1}
(\rho-\rho_{q})\nabla\rho ,
\end{eqnarray}
where $W_1$  and $\gamma$ are the two additional parameters. To
reproduce the kink of the charge radii with the increasing mass of
lead isotopes, the isospin dependence of the nuclear spin-orbit
interaction was investigated, and in Ref.~\cite{Sha95} the form
factor of the spin-orbit potential is written as
\begin{equation}
\vec{W}_{q} =\frac{W_0}{2}(1+\chi_{w})\nabla \rho_q +
\frac{W_0}{2}\nabla \rho_{q^\prime},~~(q \ne q^\prime)
\end{equation}
where $\chi_{w}$ is the parameter to mimic the isospin dependence.
For the relativistic mean-field model, the non-relativistic
expansion of the Dirac equation usually gives different density and
isospin dependence of the spin-orbit coupling compared to the
Skyrme-Hartree-Fock functional~\cite{Rei95}. In order to take the
main physics of the density and isospin dependence of the spin-orbit
coupling into consideration while making the form as simple as
possible, the form factor can be generally written as~\cite{Xu13}
\begin{equation}
\vec{W}_{q} =\frac{W_0^\star(\rho)}{2}(a\nabla \rho_q + b\nabla
\rho_{q^\prime})~~(q \ne q^\prime)
\end{equation}
with $W_0^\star(\rho)=W_0(\rho/\rho_0)^\gamma$. In the above, $W_0$,
$\gamma$, $a$, and $b$ are the corresponding parameters for the
strength, density dependence, and isospin dependence of the
spin-orbit coupling, and $\rho_0=0.16$ fm$^{-3}$ is the saturation
density. As mentioned in the introduction, the strength of the
spin-orbit coupling has been constrained between $80$ and $150$
MeVfm$^5$, while its density and isospin dependence are still under
hot debate. We will use $W_0=150$ MeVfm$^5$, $a=2$, $b=1$, and
$\gamma=0$ in the following calculations unless stated otherwise. A similar
generalization to all the spin-dependent mean-field potentials leads
to
\begin{eqnarray}
U_q^{s-even} &=& - \frac{W_0^\star(\rho)}{2} [\nabla \cdot
(a\vec{J}_q
+ b\vec{J}_{q^\prime}) ] \nonumber \\
&+& \frac{W_0^\star(\rho)}{2}(a\nabla \rho_q + b\nabla
\rho_{q^\prime})
\cdot (\vec{p} \times \vec{\sigma}) , \label{useveng}\\
U_q^{s-odd} &=& - \frac{W_0^\star(\rho)}{2}\vec{p} \cdot[\nabla
\times (a\vec{s}_q+ b\vec{s}_{q^\prime}) ] \nonumber\\
&-&\frac{W_0^\star(\rho)}{2} \vec{\sigma} \cdot [\nabla \times
(a\vec{j}_q + b\vec{j}_{q^\prime}) ]. (q \ne q^\prime)
\label{usoddg}
\end{eqnarray}

\section{Incorporating the spin degree of freedom into the IBUU transport model}
\label{sibuu12}

The isospin-dependent Boltzmann-Uehling-Uhlenbeck (IBUU) transport
model~\cite{Li97,LCK} has been very successful in studying
intermediate-energy heavy-ion collisions, especially the isospin
effects. However, in the previous studies, spin effects were
neglected as only spin-averaged quantities were studied. To
introduce spin effects into the IBUU model, each nucleon now has an
additional degree of freedom, i.e., a unit vector representing the
expectation value of its spin $\vec{\sigma}$. The spin, momentum,
and spin-current densities can be calculated by using the test
particle method~\cite{Won82,Ber88} similar to the number density
$\rho$, i.e.,
\begin{eqnarray}
\rho(\vec{r}) &=& \frac{1}{N_{test}}\sum_{\rm i} \delta(\vec{r}-\vec{r}_i),\\
\vec{s}(\vec{r}) &=& \frac{1}{N_{test}}\sum_{\rm i} \vec{\sigma}_i \delta(\vec{r}-\vec{r}_i),\\
\vec{j}(\vec{r}) &=& \frac{1}{N_{test}}\sum_{\rm i} \vec{p}_i \delta(\vec{r}-\vec{r}_i),\\
\vec{J}(\vec{r}) &=& \frac{1}{N_{test}}\sum_{\rm i} (\vec{p}_i
\times \vec{\sigma}_i) \delta(\vec{r}-\vec{r}_i),
\end{eqnarray}
where $N_{test}$ is the test particle number. In addition, the
equations of motion in the presence of the spin-dependent mean-field
potentials can now be written as
\begin{eqnarray}
\frac{d\vec{r}}{dt} &=& \frac{\vec{p}}{m} +
\frac{W_0^\star(\rho)}{2} \vec{\sigma} \times (a\nabla \rho_q +
b\nabla \rho_{q^\prime}) \nonumber \\
&-& \frac{W_0^\star(\rho)}{2} \nabla \times (a\vec{s}_q + b\vec{s}_{q^\prime}), \label{rt}\\
\frac{d\vec{p}}{dt} &=& - \nabla U_q - \nabla U_q^{s-even} - \nabla
U_q^{s-odd}, \label{pt}
\\
\frac{d\vec{\sigma}}{dt} &=& W_0^\star(\rho)[(a\nabla \rho_q +
b\nabla \rho_{q^\prime}) \times \vec{p}] \times
\vec{\sigma} \nonumber \\
&-&W_0^\star(\rho)[\nabla \times (a\vec{j}_q + b\vec{j}_{q^\prime})]
\times \vec{\sigma}. \label{sigmat}
\end{eqnarray}
As an illustration of the spin dynamics, for the spin-independent
mean-field potential $U_q$ we use a Skyrme-type momentum-independent
one, and it is fitted to give the binding energy $E_0(\rho_0)=-16$
MeV for normal nuclear matter and symmetry energy
$E_{sym}(\rho_0)=30$ MeV at saturation density. The
incompressibility $K_0$ and the slope parameter $L$ of the symmetry
energy are set to be 230 MeV and 60 MeV, respectively.

In simulating heavy-ion collisions we set $z$ as the beam axis. The
distance between the centers of the two colliding nuclei in the $x$
direction is the impact parameter b. We refer in the following a
nucleon with its spin in the $+y$ ($-y$) direction as a spin-up
(spin-down) nucleon. The spin-dependent potential thus describes the
coupling between the nucleon spin and the local angular momentum of
the system during the collision process. Since initially the spins
of nucleons are randomly distributed, the second terms in
Eqs.~(\ref{useveng}) and (\ref{usoddg}) are most important. During
the collision process, as shown in Fig.~1 of Ref.~\cite{Xu13}, the
second term in Eq.~(\ref{useveng}) ((\ref{usoddg})) is repulsive
(attractive) for spin-up nucleons and attractive (repulsive) for
spin-down ones. Within the framework of the SIBUU transport
model~\cite{Xu13} and the time-dependent Hartree-Fock
model~\cite{Mar06}, the time-odd contribution (Eq.~(\ref{usoddg}))
is larger than the time-even contribution (Eq.~(\ref{useveng})). The
net spin-dependent potential is thus more attractive (repulsive) for
spin-up (-down) nucleons. This is similar to the symmetry potential
which is repulsive (attractive) for neutrons (protons).

Besides the spin-dependent mean-field potential, the spin during
nucleon-nucleon scatterings should also be treated with care. It was
found that the spin of a nucleon may flip after nucleon-nucleon
scatterings~\cite{Ohl72} from spin-dependent nucleon-nucleon
interactions. Although it was shown that the spin-flip probability
is appreciable and dependent on the energy and momentum
transfer~\cite{Lov81}, it is still not well determined due to the
lack of accurate knowledge regarding in-medium spin-related
nucleon-nucleon interactions. In the present work, as a conservative
approximation we randomize the final spins of the two colliding
nucleons after each scattering. In addition, the nucleon phase space
distribution is further sorted according to nucleon spin to properly treat
the spin- and isospin-dependent Pauli blocking.

\section{Results and discussions}
\label{results}

In our previous work with the SIBUU12 transport model~\cite{Xu13},
we already found that the spin up-down differential transverse flow
is a useful probe of the in-medium spin-orbit coupling. Here we
study more systematically the spin-dependent transverse flow. One of
our purposes is to find the beam energy and the collision centrality at which
the spin up-down differential transverse flow is the largest and
most easily observable. We are also interested in disengaging
effects of the strength, isospin dependence, and density dependence
of the spin-orbit coupling. In our calculations, we stop the
simulations when the projectile and the target fall apart. 200,000
events are generated for each beam energy and impact parameter.

\subsection{Up-down differential transverse flow and its dependence on the spin-orbit coupling, beam energy, and impact parameter}

\begin{figure}[h]
\includegraphics[scale=0.7]{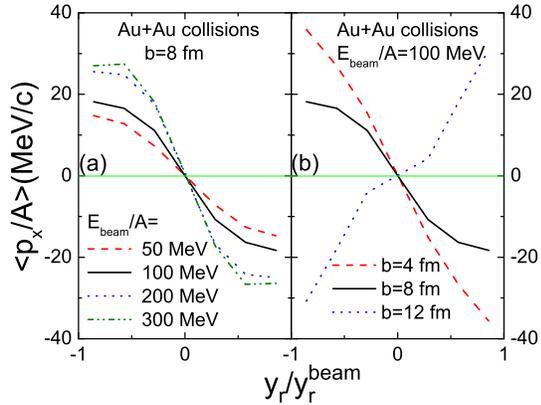}
\caption{(Color online) Transverse flow of Au+Au collisions at an
impact parameter of 8 fm and beam energies of 50, 100, 200, and 300
MeV/nucleon (left panel), and at a beam energy of 100 MeV/nucleon
and impact parameters of 4, 8, and 12 fm (right panel).}\label{F1}
\end{figure}

The nucleon transverse flow is normally revealed by examining the
nucleon average transverse momentum in the reaction plane versus
rapidity. For ease of following discussions, we first illustrate in
Fig.~\ref{F1} the spin-averaged nucleon transverse flow and its
dependence on the beam energy and impact parameter. We note that in
our code the projectile (target) is put on the $-x$ ($+x$) side in
the reaction plane. Thus, the sign of the transverse flow is
negative as shown in Ref.~\cite{Xu13}, different from the convention
that some others may have used. The main features seen here are
consistent with those found in previous studies in the
literature~\cite{Ber88,Dan02a,Dan85}. In particular, as a result of
the competition between the generally attractive mean-field
potential in the energy range considered and the repulsive
nucleon-nucleon scatterings, the transverse flow first increases
fast then slowly with increasing beam energy. At a given beam energy
of 100 MeV/nucleon, the transverse flow is largest at $\text{b}=4$
fm and changes sign at $\text{b}=12$ fm.

\begin{figure}[h]
\includegraphics[scale=0.7]{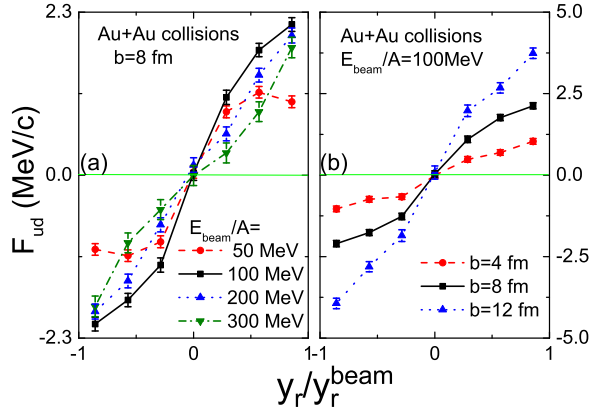}
\caption {(Color online) Same as Fig.~\ref{F1} but for spin up-down differential
transverse flow.}\label{F2}
\end{figure}

The above spin-averaged transverse flow can be altered if a
momentum-dependent spin-independent potential is used~\cite{GBD87},
while in the present study we are only interested in the spin
splitting of the transverse flow as a result of the spin-dependent
potential. As mentioned in Sec.~\ref{sibuu12}, the net
spin-dependent potential is more attractive (repulsive) for spin-up
(down) nucleons. Consequently, there is a splitting in the average
transverse momentum in the reaction plane between spin-up and
spin-down nucleons. This splitting can be measured quantitatively by
using the spin up-down differential transverse flow~\cite{Xu13}
\begin{equation}
F_{ud}(y_r) = \frac{1}{N(y_r)} \sum_{i=1}^{N(y_r)} \sigma_i (p_x)_i,
\end{equation}
where $N(y_r)$ is the number of nucleons with rapidity $y_r$, and
$\sigma_i$ is $1 (-1)$ for spin-up (spin-down) nucleons. The spin
up-down differential transverse flow maximizes the effects of the
opposite spin-dependent potentials for spin-up and spin-down nucleons
while canceling out largely spin-independent contributions. The left
panel of Fig.~\ref{F2} displays the spin up-down differential
transverse flow for Au+Au collisions at an impact parameter of 8 fm
and beam energies of 50, 100, 200, and 300 MeV/nucleon. Different
from the spin-averaged transverse flow, the slope of the
differential one first increases then decreases with increasing
collision energy, and the largest spin up-down differential
transverse flow is observed at a beam energy of about 100
MeV/nucleon. We note that the angular momentum increases with
increasing beam energy, leading to stronger spin-dependent potentials.
On the other hand, the violent nucleon-nucleon scatterings randomize
the spin direction and turn to wash out the effects from the spin-dependent
potentials, especially at higher energies. The competition of the
above effects leads to the non-monotonical behavior of the slope of
$F_{ud}$ with respect to the beam energy. In the right panel of
Fig.~\ref{F2} we show the results at a beam energy of 100
MeV/nucleon and an impact parameter of 4, 8, and 12 fm. It is
seen that the spin up-down differential flow is larger for
peripheral collisions where the statistical errors are also larger.
This is understandable as the second terms in
Eqs.~(\ref{useveng}) and (\ref{usoddg}) generally increase with the
increasing nucleon density gradient, and the latter becomes larger
near the edge of a nucleus.

\begin{figure}[h]
\includegraphics[scale=0.7]{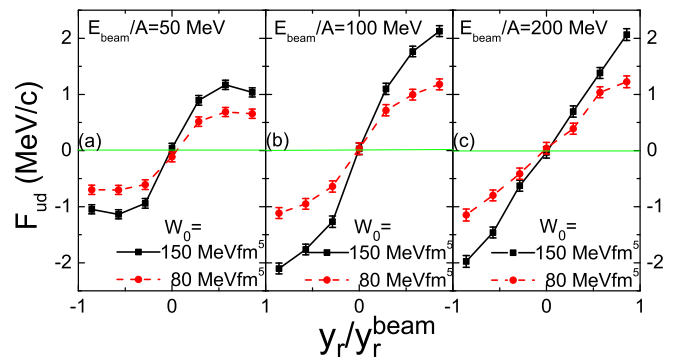}
\caption { (Color online) The spin up-down differential transverse flow with
different $W_0$ for Au +Au collisions at an impact parameter of
$\text{b}=8$ fm and beam energies of 50 (a), 100 (b), and 200 (c)
MeV/nucleon. }\label{F3}
\end{figure}

Next, we study the dependence of the spin up-down differential
transverse flow on the strength of the spin-orbit coupling $W_0$ at
different beam energies for mid-central ($\text{b}=8$ fm) Au+Au
collisions.  Shown in Fig.~\ref{F3} are comparisons of results obtained using
the upper and lower limit of $W_0$, i.e., $W_0=150$ MeVfm$^5$ and $80$ MeVfm$^5$.
To quantify the strength of the spin up-down differential flow,
we use the slope parameter $F^\prime$ of $F_{ud}$ at mid rapidity, i.e.,
\begin{equation}
F^\prime = \left[\frac{dF_{ud}}{d(y_r/y^{beam}_r)}\right]_{y_r=0},
\end{equation}
where $y_r/y^{beam}_r$ is the reduced rapidity. Similar to the
calculation of the flow parameter for the direct flow in
Ref.~\cite{Kdoss86}, we use a 3rd-order polynomial to fit the
s-shaped $F_{ud}$ curve to get its slope $F^\prime$. Moreover, to
measure quantitatively effects from varying $W_0$ we define a
sensitivity parameter as
\begin{equation}
\delta
=\frac{F_{150}^{\prime}-F_{80}^{\prime}}{F_{150}^{\prime}+F_{80}^{\prime}},
\label{del}
\end{equation}
where $F_{150}^{\prime}$ ($F_{80}^{\prime}$) is the flow parameter
$F^\prime$ from $W_0=150$ (80) MeVfm$^5$. The values of
$F_{150}^{\prime}$ and $F_{80}^{\prime}$ as well as $\delta$ at
different beam energies are listed in Table~\ref{T1}. It is seen
that the value of the sensitivity parameter $\delta$ is appreciable
and approximately a constant within the statistical errors from 50
to 200 MeV/nucleon.
\begin{table}[h]\small
  \centering
  \caption{Slope parameters for spin up-down differential transverse flow at $W_0=150$ and 80 MeVfm$^5$ as well as their relative difference $\delta$ (Eq.~(\ref{del}))
  in Au+Au collisions at an impact parameter of $\text{b}=8$ fm with different beam energies.}
    \begin{tabular}{cccc}
    \hline
    $E_{\rm beam}$ (AMeV)      & 50   & 100    & 200\\
    \hline
     $F_{150}^\prime$     & $3.60\pm 0.19$    & $4.49\pm 0.35$    & $2.25\pm 0.22$   \\
     $F_{80}^{\prime}$     & $2.23\pm 0.13$    & $2.60\pm 0.02$    & $1.33\pm 0.23$   \\
     $\delta$     & $0.23\pm 0.04$    & $0.27\pm 0.04$    & $0.26\pm 0.07$   \\
    \hline
    \end{tabular}
  \label{T1}
\end{table}

\begin{figure}[h]
\includegraphics[scale=1.2]{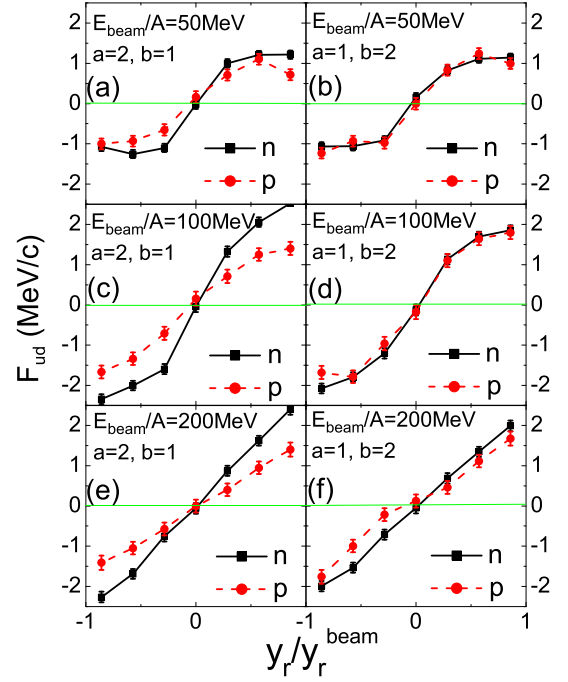}
\caption{(Color online) Spin up-down differential transverse flows for neutrons and
protons with different ratios of isospin-like and isospin-unlike
couplings, i.e., $a=2$, $b=1$ (a, c, e) and $a=1$, $b=2$ (b, d, f)
for Au+Au collisions at a beam energy of 50 (a, b), 100 (c, d), and
200 (e, f) MeV/nucleon with impact parameter $\text{b}=8$ fm.
}\label{F4}
\end{figure}

The isospin dependence of the spin-orbit coupling can be studied by
comparing the spin up-down differential transverse flow for neutrons
and protons using different strength of isospin-like and
isospin-unlike coupling, i.e., different values of $a$ and $b$ in
Eqs.~(\ref{useveng}) and (\ref{usoddg}). Here we use typical values
of ($a=2$ and $b=1$) and ($a=1$ and $b=2$), and the results for
different beam energies are shown in Fig.~\ref{F4}. We note that the
system is globally neutron rich, and the $\nabla\rho_{n}$ and
$\nabla\times \vec{j}_{n}$ are generally larger than the
$\nabla\rho_{p}$ and $\nabla\times \vec{j}_{p}$, respectively. It is
seen that with the parameter set corresponding to a stronger
isospin-like coupling, i.e., ($a=2$ and $b=1$), the $F_{ud}$ of
neutrons is larger than that of protons, while the difference of
$F_{ud}$ between neutrons and protons becomes smaller with a
stronger isospin-unlike coupling, i.e., ($a=1$ and $b=2$). To
quantify the sensitivity, we define the relative difference
$\delta^{\prime}$ as
\begin{equation}\label{delp}
\delta^{\prime}
=\frac{F_{n}^{\prime}-F_{p}^{\prime}}{F_{n}^{\prime}+F_{p}^{\prime}},
\end{equation}
where $F_{n}^{\prime}$ ($F_{p}^{\prime}$) is the value of the flow
parameter $F^\prime$ of the spin up-down differential transverse
flow for neutrons (protons). The values of $F_{n}^{\prime}$ and
$F_{p}^{\prime}$ as well as $\delta^{\prime}$ for different beam
energies in Fig.~\ref{F4} are summarized in Table~\ref{T2}. We can
see that the sensitivity is still appreciable at beam energies of 50
and 100 MeV/nucleon, as the difference of $\delta^\prime$ from the
two parameter sets are significantly larger than the statistical
error. Considering the magnitude of $F_{ud}$, the different neutron
and proton spin up-down differential transverse flows for
mid-central Au+Au collisions at the beam energy of about 100
MeV/nucleon might be the best probe of the isospin dependence of the
spin-orbit coupling.

\begin{table}[h]\tiny
  \centering
  \caption{Slope parameters for neutron and proton spin up-down differential transverse flow as well as their relative difference $\delta^\prime$ (Eq.~(\ref{delp}))
  in Au+Au collisions at impact parameter $\text{b}=8$ fm with different beam energies.}
    \begin{tabular}{c|cc|cc|cc}
    \hline
& $E_{\rm beam}=50$  & \multicolumn{1}{l|}{ (AMeV)} & $E_{\rm beam}=100$ & \multicolumn{1}{l|}{ (AMeV)} & $E_{\rm beam}=200$ & \multicolumn{1}{l}{ (AMeV)} \\
\cline{2-7}          & \multicolumn{1}{c|}{$a/b=2$} & \multicolumn{1}{c|}{$a/b=1/2$} & \multicolumn{1}{c|}{$a/b=2$} & \multicolumn{1}{c|}{$a/b=1/2$} & \multicolumn{1}{c|}{$a/b=2$} & \multicolumn{1}{c}{$a/b=1/2$} \\
    \hline
    $F_n^{\prime}$& \multicolumn{1}{c|}{$4.17\pm 0.09$} & \multicolumn{1}{c|}{$3.41\pm 0.53$} & \multicolumn{1}{c|}{$5.62\pm 0.35$} & \multicolumn{1}{c|}{$4.43\pm 0.24$} & \multicolumn{1}{c|}{$2.60\pm 0.50$} & \multicolumn{1}{c}{$2.37\pm 0.28$} \\
    $F_p^{\prime}$& \multicolumn{1}{c|}{$2.59\pm 0.36$} & \multicolumn{1}{c|}{$3.58\pm 0.34$} & \multicolumn{1}{c|}{$2.55\pm 0.33$} & \multicolumn{1}{c|}{$3.74\pm 0.75$} & \multicolumn{1}{c|}{$1.68\pm 0.23$} & \multicolumn{1}{c}{$1.10\pm 0.39$} \\
    $\delta^{\prime}$& \multicolumn{1}{c|}{$0.23\pm 0.06$} & \multicolumn{1}{c|}{$-0.02\pm 0.09$} & \multicolumn{1}{c|}{$0.38\pm 0.06$} & \multicolumn{1}{c|}{$0.08\pm 0.10$} & \multicolumn{1}{c|}{$0.21\pm 0.08$} & \multicolumn{1}{c}{$0.36\pm 0.09$} \\
    \hline
    \end{tabular}
  \label{T2}
\end{table}

\subsection{Disentangle effects of the strength and density dependence of the spin-orbit coupling with high transverse momentum nucleons}
\begin{figure}[h]
\includegraphics[scale=0.7]{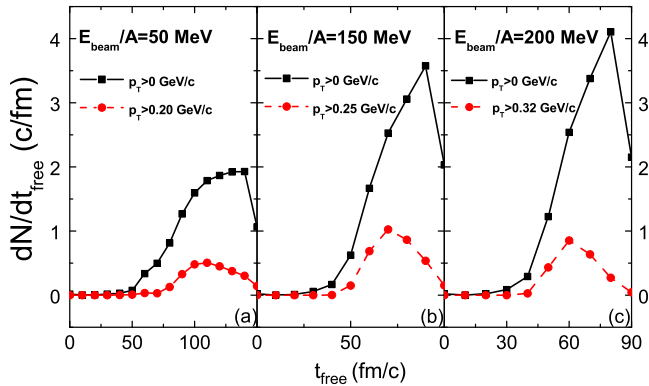}
\caption{ (Color online) Emission rates of free nucleons with and
without transverse momentum cut in Au+Au collisions at an impact
parameter of $\text{b}=8$ fm with beam energies of 50 (a), 150 (b),
and 200 (c) MeV/nucleon, respectively.}\label{F6}
\end{figure}

\begin{figure}[h]
\includegraphics[scale=0.6]{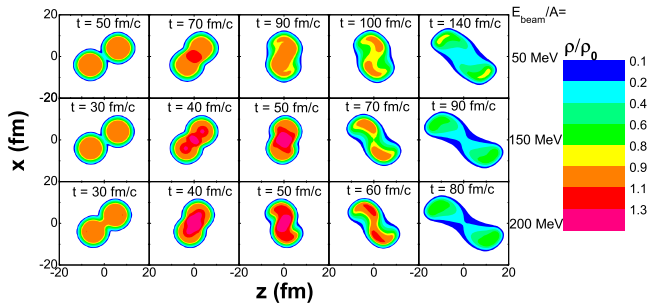}
\caption{(Color online) Contours of reduced nucleon density
$\rho/\rho_0$ in the reaction plane at different times for Au+Au
collisions at an impact parameter of $\text{b}=8$ fm and beam
energies of 50 (first row), 150 (second row), and 200 MeV/nucleon
(third row), respectively.}\label{F5}
\end{figure}

Within our framework, we use a parameterized form of the spin-orbit
coupling, i.e., $W_0(\rho/\rho_0)^\gamma$, with $W_0$ representing
the strength and $\gamma$ representing the density dependence of the
spin-orbit coupling. In our previous work~\cite{Xu13} we have shown
that the spin up-down differential transverse flow is sensitive to
both the strength and the density dependence of the spin-orbit
coupling. Thus, it is difficult to probe both $W_0$ and $\gamma$
simultaneously using a single reaction. In this subsection, we
investigate if additional information can be obtained by restricting
our analyses to high transverse momentum ($p_T$) nucleons which are
known to be mostly emitted from high density region in the earlier
stage of heavy-ion collisions. In the SIBUU12 model, a nucleon is
considered free when its local density is below $\rho_0/8$. Since
the number of high $p_T$ nucleons is very small, to have enough
statistics and ensure that our qualitative conclusions are
insensitive to the specific values of the $p_T$ cut, we analyzed a
million events for each case and used the following fiducial values
for the beam energy dependent $p_{T}$ cut: $p_T>0.20$ GeV/c for
$E_{beam}=50$ MeV/nucleon, $p_T>0.25$ GeV/c for $E_{beam}=150$
MeV/nucleon, and $p_T>0.32$ GeV/c for $E_{beam}=200$ MeV/nucleon.
Figure~\ref{F6} shows the nucleon emission rate as a function of
time with and without the high $p_{T}$ cut. It is seen that the
emission of high-$p_T$ nucleons peaks at approximately $t=100$ fm/c
for $E_{beam}=50$ MeV/nucleon, $t=70$ fm/c for $E_{beam}=150$
MeV/nucleon, and $t=60$ fm/c for $E_{beam}=200$ MeV/nucleon,
respectively. From the time evolution of the density contours shown
in Fig.~\ref{F5}, it can be seen that at these instants the
projectile and the target have just passed through each other. These
high-$p_T$ nucleons may thus carry important information about the
high-density phase of the reaction, including the time-integrated
effects of the spin-dependent mean-field potentials before their
emission.

\begin{figure}[h]
\includegraphics[scale=0.8]{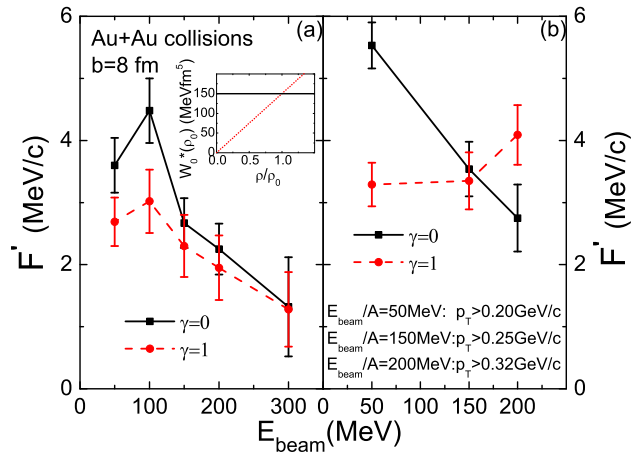}
\caption{ (Color online) The flow parameter $F^\prime$ of the spin
up-down differential transverse flow as a function of the beam
energy with $\gamma=0$ and $\gamma=1$ for all free nucleons (a) and
for high-$p_T$ free nucleons (b). The inset in panel (a) shows the
density dependence of the spin-orbit coupling
$W_0(\rho/\rho_0)^\gamma$.}\label{F7}
\end{figure}

For a comparison, we first show in Panel (a) of Fig.~\ref{F7} the
flow parameter $F^\prime$ of the spin up-down differential
transverse flow from all free nucleons as a function of the beam
energy with $\gamma=0$ and $\gamma=1$, respectively. It is seen that the
$F^\prime$ is slightly higher with $\gamma=0$ especially at lower
beam energies, as nucleons are mostly emitted from subsaturation
density regions especially at lower beam energies. Since the
spin-orbit coupling is stronger with $\gamma=0$ than $\gamma=1$ at
subsaturation densities, the $F^\prime$ is thus larger for
$\gamma=0$ than for $\gamma=1$ as already shown in Ref.~\cite{Xu13}.
However, it is clear that the observed effect is too small to be
practically useful. With the high $p_{T}$ cut, it is interesting to
see in the right panel of Fig.~\ref{F7} that the signal becomes much
stronger especially at lower beam energies. There is a cross-over in
$F^\prime$ around $E_{beam}=150$ MeV/nucleon associated with the
cross-over of the function $W_0(\rho/\rho_0)^\gamma$ shown in the
inset of Fig.~\ref{F7}. This is consistent with the observation from
Fig.~\ref{F5} that the average density where high-$p_{T}$ nucleons
are emitted is smaller (higher) than $\rho_0$ in reactions at beam
energies below (above) 150 MeV/nucleon. In this way, the strength
and the density dependence of the spin-orbit coupling are
disentangled, and they can thus be determined separately by
considering the spin up-down differential transverse flow of
high-$p_T$ nucleons at their corresponding beam energies.

\section{Summary}
\label{summary}

In summary, we have studied more systematically the spin up-down
differential transverse flow in intermediate-energy heavy-ion
collisions. Its potential for probing the in-medium spin-orbit
interaction is examined in some detail. For the Au+Au collisions
studied here, we found that the largest and most reliable spin
up-down differential transverse flow occurs in mid-central reactions
at the beam energy of about 100 MeV/nucleon. In this optimal
reaction, the strength and the isospin dependence of the spin-orbit
coupling can be extracted by measuring the slope of the differential
flow and comparing the slopes of neutrons and protons. Moreover, the
spin up-down differential transverse flow of high-$p_T$ nucleons at
different beam energies can be used to delineate the density
dependence of the spin-orbit coupling.

So far polarized rare isotope beams can already be produced with
nucleon removal or pickup reactions at RIKEN~\cite{Ich12},
GSI~\cite{Sch94}, NSCL~\cite{Gro03}, and GANIL~\cite{Tur06}, and the
spin polarization of projectile fragments can be measured through
the angular distribution of their $\beta$ or $\gamma$ decays. In
addition, at AGS or RHIC energy people are measuring the analyzing
power denoting the chiral asymmetry as well as spin flip probability
in elastic pp or pA collisions, see, e.g., \cite{AGSspin, RHICspin}.
Unfortunately, although the spin polarization of fragments are
measurable, in the current stage it is still very difficult to
measure the spin of a single nucleon experimentally. We hope our
study will stimulate the development of new facilities for
spin-related experiments so that some of the spin-dependent
observables (such as the spin-polarized light fragments) can be
measured to extract useful information of the spin-orbit coupling.
We believe that the spin-isospin physics in heavy-ion collisions at
intermediate energies will provide complementary information about
the poorly known in-medium spin-orbit force affecting many novel
features of rare isotopes.

\begin{acknowledgments}
This work was supported by the '100-talent plan' of Shanghai
Institute of Applied Physics under grant Y290061011 from the Chinese
Academy of Sciences, the National Natural Science Foundation of
China under grant No. 11320101004, the US National Science
Foundation grants PHY-1068022, the National Aeronautics and Space
Administration under grant NNX11AC41G issued through the Science
Mission Directorate, and the CUSTIPEN (China-U.S. Theory Institute
for Physics with Exotic Nuclei) under DOE grant number
DE-FG02-13ER42025.
\end{acknowledgments}

\end{document}